\documentclass[11pt, letterpaper, final]{IEEEtran}

\usepackage{amsmath,amscd,amssymb}
\usepackage{bm}
\usepackage{graphicx}
\usepackage{subfigure}
\usepackage{units}
\usepackage{array,booktabs}
\usepackage{amssymb}
\usepackage{setspace}
\usepackage{siunitx}

\newcommand{\dd}{\textrm{d}}
\newcommand{\ii}{\textrm{i}}
\renewcommand{\vec}[1]{\bm{#1}}
\newcommand{\rrr}{\vec{r}}
\newcommand{\vvv}{\vec{v}}

\begin{document}


\title{Acoustic interaction forces and torques acting on suspended spheres in an ideal fluid}
\author{J. Henrique Lopes,~\IEEEmembership{Student Member,~IEEE},  
M. Azarpeyvand and Glauber T. Silva,~\IEEEmembership{Member,~IEEE} 
\thanks{J. Henrique Lopes, Glauber T. Silva are 
with Physical Acoustics Group, Instituto de F\'isica, Universidade Federal de Alagoas, 
Macei\'o, AL 57072-900, Brazil. Corresponding author: \texttt{glauber@pq.cnpq.br} }
\thanks{M. Azarpeyvand is with  Department of Mechanical Engineering, University of Bristol, U.K.}}

\maketitle

\begin{abstract}
In this paper, the acoustic interaction forces and torques exerted by an arbitrary time-harmonic wave on a set 
of $\bm N$ spheres suspended in an inviscid fluid are theoretically analyzed. 
In so doing, we utilize the partial-wave expansion method to solve the related multiple scattering problem.
The acoustic interaction force and torque are computed for a sphere using
the  farfield radiation force and torque formulas.
To exemplify the method, we calculate the interaction forces exerted by an external traveling and standing plane wave
 on an arrangement of two and three olive-oil droplets in water.
 The droplets radii are comparable
 to the wavelength (i.e. Mie scattering regime). 
The results  show that the radiation force may considerably deviates from that exerted solely by the external incident 
wave.
In addition, we find that acoustic interaction torques arise on the droplets when a
nonsymmetric effective incident wave interacts with the droplets.
\end{abstract}


\section{INTRODUCTION}
\IEEEPARstart{T}he possibility of noncontact particle handling by means of the acoustic radiation force (also known as acoustophoresis)
is burgeoning in microbiology and biotechnology applications~\cite{laurell}-\cite{li}.	
One of the first experimental steps in acoustophoresis was the pioneering work by Whymark~\cite{whymark},
who developed an acoustic resonant chamber for noncontact material positioning.
The field gained another important contribution with the development of acoustical tweezers based on focused ultrasound beams~\cite{wu}.
New opportunities for acoustophoresis were opened with the fabrication of microfluidics devices~\cite{evander} and \cite{jones}.
The increasing number of different applications based on acoustophoresis has boosted 
an interest on understanding and proper utilization of the acoustic radiation force.
  
The phenomenon of  acoustic radiation force is  caused by the transferring of momentum flux from an incident wave to 
a suspended object ~\cite{torr}. 
The radiation force exerted on a single sphere has been subject to extensive investigation~\cite{king}-\cite{silva2014a}. 
Moreover, the angular momentum of a wave with respect to the center of a sphere can produce a 
torque on the sphere, which has also been the subject of much research~\cite{maidanik}-\cite{zhang2013}. 

In most applications of acoustic particle manipulation, an external ultrasound wave interacts with several objects suspended in a fluid. 
Thus, understanding how the acoustic radiation force is generated on multiple objects is a crucial step for development 
and improvement of acoustic manipulation methods.

The acoustic interaction forces for a system of two or more spheres has been 
previously studied. 
The mutual force between two spheres with the axial connection line parallel to the propagation direction  of 
a plane wave was studied by Emblenton~\cite{embleton}. 
Crum~\cite{crum} analyzed the mutual interaction forces of two bubbles 
in a stationary sound field. 
Investigations of the interaction forces at small separation distances for different particle pairs 
(bubble-solid, bubble-drop, solid-solid) have been carried out by Doinikov \cite{doinikov3}-\cite{doinikov4}. 
Apfel~\cite{apfel} computed the acoustic radiation force on two fluid spheres under a plane wave. 
Doinikov~\cite{doinikov2} presented an analytical expression for the acoustic radiation force 
on a $N$-sphere system,  obtained by integrating the acoustic momentum flux on the surface
of each particle (nearfield method). 
Recently, Azarpeyvand \emph{et al}.~\cite{mahdi2013} computed the axial radiation force resultant from a 
Bessel beam on an acoustically 
reflective sphere in the presence of an adjacent 
spherical particle immersed in an inviscid fluid. 
 Silva \emph{et al}.~\cite{silva2014} analyzed the  acoustic interaction forces
in a system of many suspended particles in the Rayleigh scattering limit.
On the other hand, the acoustic interaction torques in a many-body system
has not been analyzed before.

In this work, we present  a method to compute the acoustic interaction forces
and torques resulting from the interaction by suspended spheres and an external ultrasound field of arbitrary wavefront.
In other words, the induced radiation force and torque is obtained for each sphere in the medium.
To do so, the effective incident acoustic wave on each sphere should be
calculated by solving the corresponding multiple-body scattering problem.
Due to the symmetry of the spherical objects, it is convenient to formulate the many-body
scattering problem using the partial-wave expansion in spherical coordinates~\cite{gaunaurd}.
The complete solution of the many-body scattering problem requires that all partial-wave expansions 
 be expressed in the same coordinate system.
This is accomplished by means of the additional theorem for spherical functions~\cite{ivanov},
which relates the partial-waves between two arbitrary coordinate systems.
In this manner, the effective incident wave on each sphere and its corresponding scattered wave field
are determined by numerically solving a system of linear equations, involving the partial-wave expansion 
coefficients.
Upon obtaining these coefficients, one can then determine the radiation forces and torques exerted on the spheres
 using the expressions developed in Refs.~\cite{silva:epl}, \cite{silva:3541}, and \cite{silva1207}. 
The method is applied to arrangements of two and three olive oil droplets.
The results reveal that the acoustic interaction  forces exerted on the droplets may
remarkably deviate from the radiation force obtained for noninteracting droplets (i.e. without re-scattered waves).
In addition, the acoustic interaction torque may arise on the droplets even if 
the external wave does not  possess angular momentum.

\section{Model assumptions}
Consider an inviscid fluid of infinite extent characterized by  ambient density $\rho_0$ and  speed of sound $c_0$.
A time-harmonic wave of angular frequency $\omega$ is described by 
a velocity potential function $\phi(\rrr) e^{-\ii \omega t}$, where
$ \rrr$ denotes  position vector with respect to a system $O$ and $t$ is time.
In Cartesian coordinates, position vector is
$\bm{r}= r (\sin \theta\cos\varphi \bm{e}_x +\sin \theta\sin\varphi \bm{e}_y +\cos \theta \bm{e}_z)$,
where $\bm{e}_i$ $(i \in \{x,y,z\})$ is the Cartesian unit-vector, 
$\theta$ and $\varphi$ are the polar and azimuthal angles in spherical coordinates.

We restrict our analysis to low-amplitude waves, whose pressure $p$  obeys the condition
$|p|/(\rho_0 c_0^2)\ll 1$.
In this case, the velocity potential amplitude satisfies the Helmholtz equation
\begin{equation}
(\nabla^2 + k^2)\phi(\rrr) = 0,
\end{equation}
where $k=\omega/c_0$ is the wavenumber.
The time-dependent term $e^{-\ii \omega t}$  is omitted for the sake of simplicity.
The amplitude of the excess of pressure and  fluid velocity are given in terms of the potential
 function $\phi$, respectively, by
\begin{align}
\label{pressure}
p(\rrr) &= -\ii \rho_0 c_0 k\phi(\rrr),\\
\vvv(\rrr)&= - \nabla \phi(\rrr). 
\label{velocity}
\end{align}

Assume that an \emph{external} incident wave interacts with a set of $N$ spheres $(N\ge 2)$.
The spheres have have radii $a_q$ ($q=1,2,\dots,N$), density $\rho_q$, and speed of sound $c_q$.
They  are arbitrary located at $\rrr'_q$  with respect to $O$  (see Fig. 1).
The center of each sphere defines a coordinate system denoted by $O_q$.
One of the spheres is referred to as \emph{probe}  ($q=p$),
while the others are \emph{sources}.

Due to the transfer of the linear and angular momentum from the external wave,
an external radiation force and torque may appear on the spheres.
Moreover, the multiple-scattered waves between the sphere may give rise to time-averaged
acoustic interaction forces and torques  between the spheres.
To calculate the acoustic interaction force and torque exerted on the probe sphere, for instance,
we have to compute the effective incident wave.
This wave  results from the combination
of the external and re-scattered waves by all source spheres in the medium.
\begin{figure}
\centering
\includegraphics[scale=.45]{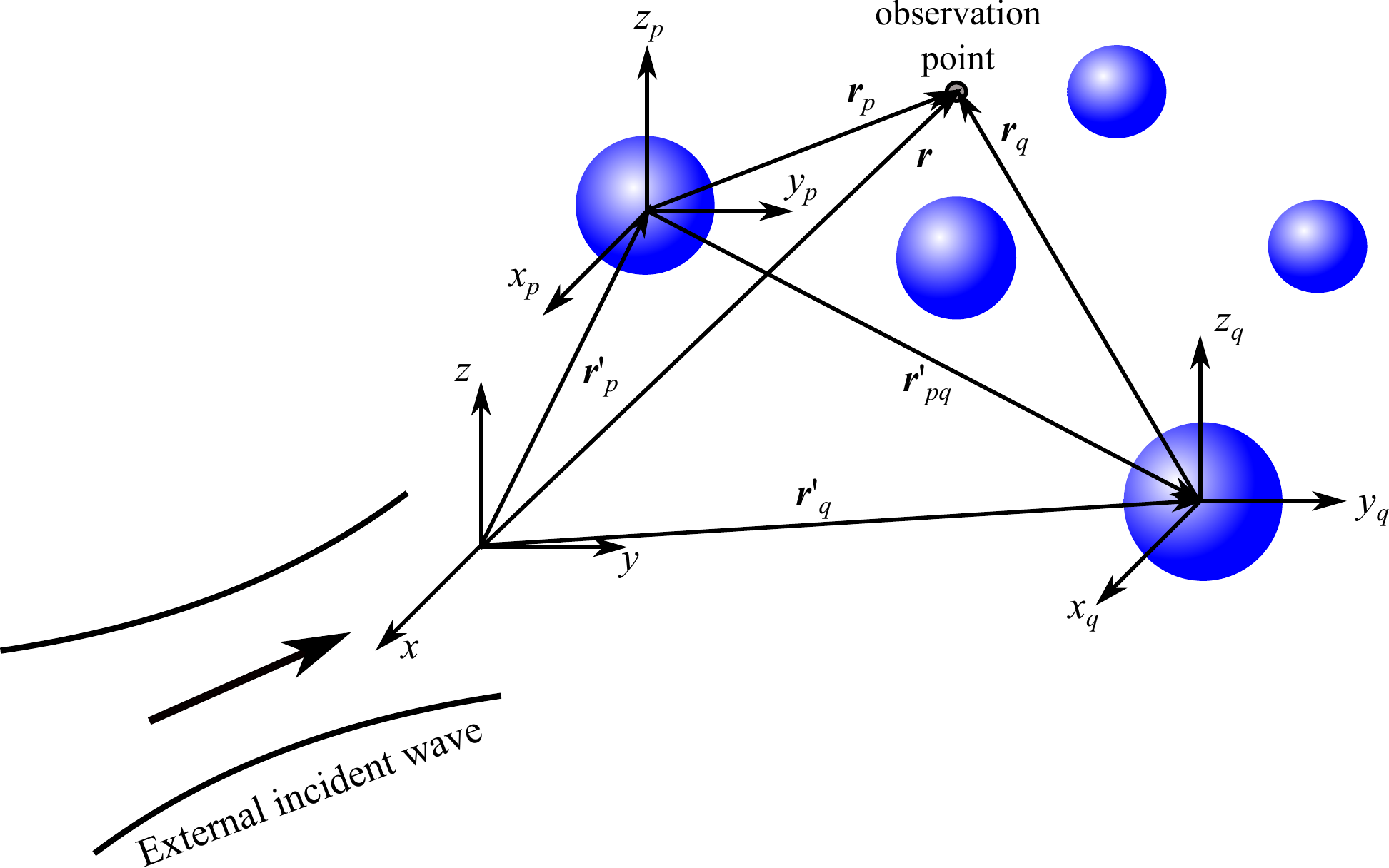}
\caption{Geometric description of the multiple scattering by particles in a suspension. 
A source particle is located at $\rrr'_{q}$ and its corresponding scattered wave is probed
at  $\rrr'_{p}$.
The observation point is located at $\rrr$, $\rrr_p$, $\rrr_q$ with respect to systems $O$, $O_p$, and $O_q$.
\label{fig:geometry}}
\end{figure}


\section{Multiple scattering}
The potential amplitude of the external incident wave can be expanded in  spherical partial-wave series 
with respect to the system $O_p$, which is located at the probe sphere center. 
Thus, we express the normalized external velocity potential as
\begin{equation}
\label{phi_ext}
 \phi_{\text{ex},p}(\rrr_p) = \sum_{n,m} a_{nm,p} J_{n}^{m}(k\rrr_p),
\end{equation}
where $\sum_{n,m} = \sum_{n=0}^{\infty}\sum_{m=-n}^{n}$,
$a_{nm,p}$ $(n\ge 0, -n\le m \le n)$ are the expansion coefficients referred to as the external beam-shape coefficient,
and $\rrr_p$ is position vector in the system $O_p$.
The regular partial-wave is given by
\begin{equation}
 J_n^m(k\rrr_p) = j_n( k r_p) Y_n^{m}(\theta_p,\varphi_p),
\end{equation}
where
$j_n$ is the spherical Bessel function of order $n$ and
$Y_n^{m}$ is the spherical harmonic of $n$th-order and $m$th-degree.
It is worth mentioning that the beam-shape coefficients depend on the choice of the coordinate system.
These coefficients can be obtained with respect to other coordinate systems, say $O_q$ ($q\neq p$), using the
translational addition theorem for spherical functions. 
In the Appendix, we derive an expression relating the beam-shape coefficients from $O_p$ and $O_q$ systems.
According to (\ref{app:bs_sc}), we have
\begin{equation}
\label{anm_q}
 a_{nm,q} =\sum_{\nu,\mu} a_{\nu \mu, p} S_{n\nu}^{m\mu,1}(\rrr_{pq}'), \quad q\neq p,
\end{equation}
where $S_{n\nu}^{m\mu,1}(\rrr_{pq}')$ is the translation coefficient of first-type given in (\ref{addition1}) and
 $\rrr_{qp}' = \rrr_{q}'- \rrr_{p}' = \rrr_{q}- \rrr_{p}$ (see Fig.~\ref{fig:geometry}).

The velocity potential function for the scattered field by the probe sphere is given by
\begin{equation}
 \phi_{\text{sc},p}(\rrr_p) = \sum_{n,m}s_{nm,p} H_{n}^{m}(k\rrr_p),
 \label{phi_sca}  
\end{equation}
where $s_{nm,p}$ is the scattering coefficient with respect to $O_p$, which will 
be obtained using the boundary conditions on the probe sphere surface.
The scattered partial-wave is 
\begin{equation}
 H_{n}^{m}(k\rrr_p)= h_n^{(1)}(k r_p) Y_n^{m}(\theta_p,\varphi_p),
\end{equation}
where $h_n^{(1)}$ is the spherical Hankel function of first-kind.
Note that (\ref{phi_sca}) satisfies the Sommerfeld radiation condition.

The external incident wave can penetrate into the probe sphere. 
Thus, the transmitted velocity potential is
\begin{equation}
 \label{phi_tra}
 \phi_{\text{tr},p}(\rrr_p) = \sum_{n,m} t_{nm,p} J_{n}^{m}(k_{p}\rrr_p),
\end{equation}
where $k_p=\omega/c_p$ and $t_{nm,p}$ are the transmission coefficients to be determined using the appropriate boundary conditions on
the probe sphere surface. 

Since we want to compute the radiation force and torque on the probe sphere, 
we need to know the total acoustic field resulting from all the other source spheres
evaluated with respect to the system $O_p$.
To do so, we first notice that the scattered sound field by a sphere located at $\bm{r}'_q$  can be expressed in  $O_p$
as
\begin{equation}
\label{phi_sca_qp}
\phi_{\text{sc},qp}(\rrr_p) = \sum_{n,m} s_{nm,qp} J_n^m(k \rrr_p),
\end{equation}
where $s_{nm,qp}$ are the source-probe scattering coefficient with respect to $O_p$.

To determine the source-probe scattering coefficients,
we use the translational addition theorem for spherical functions.
According to (\ref{app:bs_sc}), we find
\begin{equation}
\label{snm_qp}
s_{nm,qp} = \sum_{\nu,\mu} s_{nm,q}S_{n\nu}^{m\mu,2} (\rrr_{qp}'),
\end{equation}
where $s_{nm,q}$ are the scattering coefficients in (\ref{phi_sca}) with $p=q$, and
$S_{n\nu}^{m\mu,2} $ is the translation coefficient of second-type which is given in (\ref{addition1}).

We can express the total velocity potential function outside the probe sphere as
\begin{equation}
\label{phi_rp}
 \phi_p(\rrr_p) =  \phi_{\text{in},p}(\rrr_p)+ \phi_{\text{sc},p}(\rrr_p),
\end{equation}
where the  effective incident velocity potential  to the probe sphere  is given by
\begin{equation}
 \label{eff_potential}
 \phi_{\textrm{in},p}(\bm{r}_p)
 =  \phi_{\textrm{ex},p}(\bm{r}_p) + \sideset{}{'}\sum_{q=1}^{N} \phi_{\textrm{sc},qp}(\bm{r}_p),
\end{equation}
where the primed sum means $q\neq p$.
The boundary conditions on the surface of the probe sphere are 
the continuity of the pressure and radial component of the fluid velocity.
Therefore, from (\ref{phi_rp}) we obtain the following boundary conditions at $r_p=a_p$,
\begin{align}
\label{bc1}
\phi_{\text{ex},p} + 
 \phi_{\text{sc},p} +
\sideset{}{'} \sum_{q=1}^{N} \phi_{\text{sca},qp} &= \phi_{\text{tr},p},\\
 \partial_{r_p} \phi_{\text{ex},p} + 
 \partial_{r_p}\phi_{\text{sc},p} +
 \sideset{}{'} \sum_{q=1}^{N} \partial_{r_p}\phi_{\text{sc},qp} &= \partial_{r_p} \phi_{\text{tr},p}.
\label{bc2}
 \end{align}
 Here, we used the shorthand notation for the derivative $\partial_{r_p}=\partial/\partial {r_p}$.
 Likewise, this can be repeated for any other spheres.
 Thus, using (\ref{phi_ext}), (\ref{phi_sca}), (\ref{phi_tra}), (\ref{phi_sca_qp}),
and (\ref{snm_qp}) into (\ref{bc1}) and (\ref{bc2}), 
one finds that the scattering coefficient for all suspended spheres is given by
\begin{equation}
\label{sp_nm}
s_{nm,q} = s_{n,q}b_{nm,q}, \quad q=1,2,\dots,N,
\end{equation}
where $s_{n,q}$ are the scaled coefficients
to be determined later, and
$b_{nm,q}$ are the  beam-shape coefficient  that represent the compound
wave formed by superposition of the external wave and all other scattered waves as given in (\ref{eff_potential}).
Explicitly, the effective incident wave to a sphere located at $\rrr'_q$ is expanded as follows
\begin{equation}
\label{phi_in}
 \phi_{\text{in},q}(\rrr_q) = \sum_{n,m} b_{nm,q} J_{n}^{m}(k\rrr_q).
\end{equation}
Substituting this result into (\ref{eff_potential}) along with (\ref{phi_ext}), (\ref{phi_sca_qp}), and (\ref{snm_qp}),
we find that the effective beam-shape coefficients 
  satisfy a system of linear equations, 
\begin{equation}
b_{nm,q} -  \sideset{}{'}\sum_{l=1}^{N}\sum_{\nu,\mu} s_{\nu, l}  S_{n\nu}^{m\mu,2} 
(\rrr_{l p}') b_{\nu\mu, l} = a_{nm,q},
\label{bnm2}
\end{equation}
where  the primed sum means that $l\neq q$.
The scaled scattering coefficient reads
\begin{align}
\nonumber
s_{n,q} &=  \det
\left[
\begin{matrix}
 \gamma_q j_n(k a_q) & j_n(k_q a_q)\\
j_n'(k a_q) & j_n'(k_q a_q)
\end{matrix}
\right] \\
&\times \det
\left[
\begin{matrix}
 -\gamma_q h_n(k a_q) & j_n(k_q a_q)\\
-{h_n}'(k a_q) & j_n'(k_q a_q)
\end{matrix}
\right]^{-1},  
\label{sn}
\end{align}
where $\gamma_q= k_q \rho_0 /k \rho_q $ and $k_q=\omega/c_q$ is the wavenumber inside the sphere
place at $\rrr_q'$.
An absorbing fluid sphere can be accounted for by introducing a complex wavenumber in the form~\cite{szabo}
\begin{equation}
k_{q}= \frac{\omega}{c_{q}}+\text{i}\alpha_q,
\end{equation}   
where $\alpha_q = \alpha_{0,q} (\omega/2\pi)^2$, with $\alpha_{0,q}$ being the sphere absorption coefficient.

The system of linear equations in (\ref{bnm2}) 
has an infinite number of unknown variables.
To  solve (\ref{bnm2}) we need to impose a truncation $n=M$ in the number of modes entering the calculation.
After imposing the truncation, the system of linear equations in Eq. (\ref{bnm2}) can be represented in a matrix format~\cite{gaunaurd,hasheminejad} as ${\bf A} {\bm x} = {\bm y}$,
where ${\bf A}$ is the coefficients matrix of dimension $N(M+1)^2 \times N(M+1)^2$,
${\bm x} $ and ${\bm y}$ are respectively, the column vectors of the unknowns and the  beam-shape
coefficients $a_{nm,q}$  and of dimension $1\times N(M+1)^2$.
Hence, the solution of (\ref{bnm2}) can be expressed as ${\bm x} ={\bf A}^{-1} {\bm y}$.
Once the effective beam-shape  coefficients $b_{nm,q}$ are determined, 
the acoustic radiation force and torque can be computed, as discussed in the next section.
In calculating the unknown scattering coefficients, $\bm{x}$, special care must be taken because
 the above mentioned matrix system may become ill-conditioned at high frequencies or when the particles 
 are very close to one another, producing numerical errors. 
 This has been discussed in other articles on the use of addition theorem ~\cite{gaunaurd} and \cite{hasheminejad}-\cite{hasheminejad6}. 

The truncation $n=M$ in the system of linear equation in (\ref{bnm2}) yields an approximate solution
for the effective incident potential to a probe sphere placed at $\rrr_p'$.
Thus, the potential expansion of the $M$th-order  incident  and its corresponding scattered wave are
\begin{align}
\label{phi_in_M}
\phi^{(M)}_{\text{in},p}(\rrr_p) &= \sum_{n,m}^M  b_{nm,p} J_{n}^{m}(k\rrr_p),\\
\phi^{(M)}_{\text{sc},p}(\rrr_p) &= \sum_{n,m}^M  s_{n,q} b_{nm,p} H_{n}^{m}(k\rrr_p),
\label{phi_sc_M}
\end{align}
 where $\sum_{n,m}^M =\sum_{n=0}^M \sum_{m=-n}^n$.
 Note that $b_{nm,p} = 0$ for $n>M$.
 Furthermore, the effective incident and scattered wave of the probe sphere are given by
 \begin{equation}
 \phi_{\text{in},p}= \lim_{M\rightarrow \infty}  \phi_{\text{in},p}^{(M)}\quad \textrm{and}\quad \phi_{\text{sc},p}= \lim_{M\rightarrow \infty}  \phi_{\text{sc},p}^{(M)}.
 \label{limit_phi}
 \end{equation}
 
\section{Acoustic interaction force}
The linear momentum transferring and stresses acting on an object
give rise to a time-averaged force referred to as acoustic radiation force.
To calculate this force, consider the momentum conservation equation of an ideal fluid~\cite{westervelt:1951}
\begin{equation}
\label{eq_momentum}
\partial_t(\rho \bm{v}) + \nabla \cdot {\bf S} = \bm{0},
\end{equation}
where ${\bf S} =  p {\bf I} + \rho \bm{v}\bm{v}$ is the stress tensor, with $\rho$ and $\rho \bm{v}\bm{v}$
being the fluid density and and momentum flux, respectively.
The quantity $\bf I$ is the $3\times 3$-unit matrix.
On taking the time-average over one wave cycle of (\ref{eq_momentum}), we obtain
\begin{equation}
\nabla \cdot \overline{\bf S} = \bm{0},
\end{equation}
where $\overline{\bf S}$ is the radiation stress tensor and the overbar denotes time-average.
So far, we have considered low-amplitude waves. 
Thus, we assume that the radiation stress can be expressed in second-order approximation of the pressure
amplitude as follows~\cite{westervelt:1951}
\begin{equation}
\overline{\bf S}= -\left(\frac{\rho_0|v|^2}{4}- \frac{|p|^2}{4\rho_0 c_0^2} \right){\bf I}
+\textrm{Re}\left[ \frac{1}{2}\rho_0 \bm{v}\bm{v}^* \right],
\end{equation}
where `Re' is the real-part and asterix means complex conjugation.

Consider an object of surface  $\partial \Omega$ and volume $\Omega$.
The  radiation force induced on the object by an acoustic wave is defined as
\begin{equation}
\bm{F} = \int_{\partial \Omega} \overline{\bf S} \cdot \bm{n}\: \dd^2 \bm{r},
\end{equation}
where $\bm{n}$ is the object's normal unit-vector pointing outwardly and $\dd^2 \bm{r}$ is the surface element. 

We define a control spherical surface $\partial \Omega_\textrm{c}$ of radius $R$ and volume $\Omega_\text{c}$, 
which encloses the object.
Using the Gauss divergence theorem we find
\begin{equation}
\int_{\Omega_\text{c}-\Omega} \nabla \cdot \overline{\bm{S}} \: \dd^3\rrr 
=-\int_{\partial \Omega} \overline{\bf S} \cdot \bm{n} \: \dd^2 \bm{r}+
\int_{\partial \Omega_\text{c}} \overline{\bf S} \cdot \bm{n}_\text{c} \:\dd^2 \bm{r}=0,
\end{equation}
where $\bm{n}_\text{c}$ is the outward unit-vector along the normal  to $\partial \Omega_\text{c}$.
If we take the farfield limit for the control surface $kR\gg1$, the radiation force on the object becomes
\begin{equation}
\label{RF_farfield}
\bm{F}= R^2
\int_{4 \pi} \overline{\bf S} \cdot \bm{e}_r \:\dd \Omega,
\end{equation}
where $\bm{e}_r$ is the radial unit-vetor and $\dd\Omega$  is the differential solid angle.
Therefore, the radiation force on  a single object  can be evaluated in the farfield $kR\gg1$ for which the acoustic
fields are represented by simpler spherical Bessel and Hankel functions as explained in Ref.~\cite{silva1207}.
\begin{figure*}
\begin{align}
\label{yxyy}
 F_{x,p}^{(M)} + \text{i}F_{y,p}^{(M)} &= \frac{\ii E_0 }{2k^2} \sum_{n,m}^M
 \sqrt{\frac{(n+m+1)(n+m+2)}{(2n+1)(2n+3)}}
\bigl[ S_{n,p} b_{nm,p} b_{n+1,m+1,p}^{*}  + S_{n,p}^{*} 
 b_{n,-m,p}^{*} b_{n+1,-m-1,p}  \bigr], \\
 F_{z,p}^{(M)} &= \frac{E_0}{k^2}\textrm{Im}\biggl[ \sum_{n,m}^M 
 \sqrt{\frac{(n-m+1)(n+m+1)}{(2n+1)(2n+3)} } S_{n,p} b_{nm,p} b_{n+1,m,p}^{*} \biggr],
\label{Yxyz}
\end{align}
\end{figure*}

The farfield method used to obtain the acoustic radiation force described in (\ref{RF_farfield})  
cannot be used straightforwardly to the many-body problem involving $N$ spheres.
To see this, consider that each sphere in the suspension has surface denoted by 
$\partial \Omega_q$ $(q=1,2,\dots,N)$.
The control sphere of radius $R$ encloses all suspended spheres in the medium.
Moreover, let $\bm{F}_q$ denote the radiation force exerted on a sphere placed at $\rrr_q'$.
In this case, the radiation stress tensor involves the external and the total scattered fields.
Hence, the farfield method yields
\begin{equation}
\int_{\partial \Omega_\text{c}} \overline{\bf S} \cdot \bm{n}_\text{c} \:\dd^2 \bm{r}
-\sum_{q=1}^N\int_{\partial \Omega_q} \overline{\bf S} \cdot \bm{n}_q \: \dd^2 \bm{r}_q
=0.
\end{equation}
Therefore,
\begin{equation}
\sum_{q=1}^N\bm{F}_q = R^2\int_{4 \pi}  \overline{\bf S} \cdot \bm{e}_r \: \dd \Omega.
\label{RF_farfield_many_body}
\end{equation}
With this scheme we obtain the total radiation force exerted by the external beam on the spheres.
Clearly, if we want to benefit from the farfield radiation force method, 
a new approach to the many-body problem is on demand.

To compute the radiation force with the farfield method,
assume that radiation stress on the probe sphere due to the effective incident and scattered waves
is represented by $\overline{{\bf S}_p}$.
On the other hand, the $M$th-order radiation stress tensor $\overline{{\bf S}^{(M)}_p}$, which is related to
the $M$th-order solutions for the effective incident and scattered waves in (\ref{phi_in_M}) and (\ref{phi_sc_M}),
is given by
\begin{align}
\nonumber
\overline{{\bf S}_p^{(M)}}&= \left(\frac{\rho_0\left|v^{(M)}_p\right|^2}{4}
- \frac{\left|p^{(M)}_p\right|^2}{4\rho_0 c_0^2} \right){\bf I}\\
&+\textrm{Re}\left[ \frac{1}{2}\rho_0 \bm{v}^{(M)}_p\bm{v}^{(M)*}_p \right],
\end{align}
where $p_p^{(M)} = p_{\textrm{in},p}^{(M)} + p_{\textrm{sc},p}^{(M)}$ is
the total pressure
and $\bm{v}_p^{(M)} = \bm{v}_{\textrm{in},p}^{(M)} + \bm{v}_{\textrm{sc},p}^{(M)}$
is the total fluid element velocity.
The pressures and fluid velocities should be calculated from (\ref{phi_in_M}) and (\ref{phi_sc_M})
using (\ref{pressure}) and (\ref{velocity}), respectively.
Moreover, we notice that 
\begin{equation}
\label{divS}
\nabla_p \cdot \overline{{\bf S}^{(M)}_p} = \bm{0},
\end{equation}
where $\nabla_p$ is the gradient operator in $O_p$.
Furthermore, from (\ref{limit_phi}) we have
\begin{equation}
\overline{{\bf S}_p} = \lim_{M\rightarrow \infty} \overline{{\bf S}^{(M)}_p}.
\end{equation}

Now referring to the  radiation force in  (\ref{RF_farfield}), we may express the approximate radiation force
caused by the external and the $M$th-order re-scattered waves on the probe sphere as
\begin{equation}
\bm{F}^{(M)}_p = - R^2\int_{4 \pi}  \overline{{\bf S}^{(M)}_p} \cdot \bm{e}_{r_p}  \dd \Omega_p,
\end{equation}
where $R$ is the radius of the control sphere centered at $O_p$, $\bm{e}_{r_p}$ is the radial unit-vector 
 in $O_p$, and $\dd \Omega_p$ is the differential solid angle in $O_p$.
 We can obtain the Cartesian components of the radiation force $\bm{F}_p$ 
 following the procedure described in Refs.~\cite{silva1207,silva:3541}.
The result is given in (\ref{yxyy}) and (\ref{Yxyz}), 
where $E_0=p_{0}^{2}/(2\rho_{0}c_{0}^{2})$ is the characteristic energy density of the incident wave,
with $p_0$ is the pressure magnitude,  `Im' denotes the imaginary-part, and
\begin{equation}
\label{Sn}
S_{n,p} = s_{n,p}+ s_{n+1,p}^{*} + 2 s_{n,p} s_{n+1,p}^{*}.
\end{equation}

The acoustic radiation force due to the external wave  on the probe sphere $\bm{F}_{\textrm{ex},p}$
is obtained from (\ref{yxyy}) and (\ref{Yxyz}) by replacing the  beam-shape coefficient
$b_{nm,p}$ by  $a_{nm,p}$ and setting $M\rightarrow \infty$.
Finally, the acoustic interaction force between the probe sphere and all other spheres in the medium is
obtained through the expression
\begin{equation}
 \vec{F}^{(M)}_{\text{int},p} = \vec{F}^{(M)}_p - \vec{F}_{\text{ex},p}.
\end{equation}

\begin{figure*}
\begin{align}
\nonumber
N_{x,p}^{(M)}  + \text{i}N_{y,p}^{(M)}  &=-\frac{E_0}{2 k^3}
\sum_{n,m}^M\sqrt{(n-m)(n+m+1)}\\
 \label{tauxy}
 &\times \biggl[\left(1 + s_{n,p}\right)s_{n,p}^* b_{nm,p} b_{n,m+1,p}^{*} 
 + \left(1 + s_{n,p}^*\right)s_{n,p}  b_{n,-m,p}^{*}  b_{n,-m-1,p}  \biggr],  \\
N_{z,p}^{(M)} &= -\frac{E_0}{k^3} \text{Re} \left[\sum_{n,m}^M 
m  (1 + s_{n,p})s_{n,p}^{*} \left|b_{nm,p}  \right|^2\right].
\label{tauz}
\end{align}
\end{figure*}

\section{Acoustic interaction torque}
A incident wave   may 
produce a time-averaged torque on an object of surface $\partial \Omega$ and volume $\Omega$, 
with respect to an axis passing through the object.
This effect is known as the acoustic radiation torque.
With respect to the system $O$, the acoustic radiation  torque is defined as~\cite{maidanik}
\begin{equation}
\bm{N} = \int_{\partial \Omega} \left(\bm{r} \times \overline{\bf S} \right)\cdot \bm{n} \: \dd^2 \bm{r},
\end{equation}
where $\bm{n}$ is the outward unit normal vector of $\partial \Omega$.
The quantity $\bm{r} \times \overline{\bf S}$ is the angular momentum of the radiation stress. 
By taking the vector product  $\bm{r} \times$ of (\ref{eq_momentum}),
and performing a time-average in the wave cycle, we obtain
\begin{equation}
\nabla \cdot \left(\bm{r} \times \overline{\bf S} \right) = \bm{0}.
\end{equation}
Thus, the angular momentum of the radiation stress is also a divergenceless quantity.
It is possible then to calculate the radiation torque with the acoustic fields evaluated on a spherical control surface
of radius $R$,
which encloses the object, at the farfield $kR\gg 1$.
Similarly to the radiation force on a single object given in (\ref{RF_farfield}),
the radiation torque is given by
\begin{equation}
\bm{N} = \int_{4\pi} \left(\bm{r} \times \overline{\bf S} \right)\cdot \bm{e}_r \: \dd \Omega, \quad kR\gg 1.
\end{equation}

We have demonstrated that we cannot use the farfield method with the radiation stress $\overline{\bf S}$ to calculate 
the radiation force problem involving a $N$-sphere suspension in (\ref{RF_farfield_many_body}).
Similar limitation is found in the calculation of the radiation torque on each sphere of the suspension.
Nevertheless, from (\ref{divS}), we have that the $M$th-order radiation stress related to a probe sphere satisfies
\begin{equation}
\nabla_p \cdot \left[\bm{r}_p \times \overline{ {\bf S}_p^{(M)}} \right] = \bm{0}.
\label{div_xS}
\end{equation}
Here the angular momentum of the radiation stress is defined with respect to the probe sphere position 
 that defines $O_p$.
 
Using the Gauss' divergence theorem and (\ref{div_xS}), one can show that the $M$th-order
radiation torque on the probe sphere is given by
\begin{equation}
\vec{N}_{p}^{(M)} =R^2 \int_{4\pi} \left[\rrr_{p} 
\times \overline{\mathbf{S}^{(M)}_p}\right]\cdot \vec{e}_{r_p} \:\dd\Omega_{p}, \quad kR \gg 1.
\label{Np}
\end{equation}
From this equation we can obtain the Cartesian components of the radiation torque as developed in Ref.~\cite{silva:epl}.
The result is shown in (\ref{tauxy}) and (\ref{tauz}).

If the external incident wave has angular momentum with respect to $O_p$,
the radiation torque  $\bm{N}_{\textrm{ex},p}$ can be obtained 
by setting $b_{nm,p}=a_{nm,p}$ in (\ref{tauxy}) and (\ref{tauz}) and making $M\rightarrow \infty$.
Moreover, the acoustic interaction torque between the probe with the source spheres is
given by
\begin{equation}
\bm{N}_{\textrm{int},p}^{(M)} = \bm{N}_{p}^{(M)} - \bm{N}_{\textrm{ex},p}.
\end{equation}
However, if the wave does not have angular momentum then $\bm{N}_{\textrm{ex},p}=\bm{0}$.
Consequently, the acoustic interaction torque is 
$\bm{N}_{\textrm{int},p}^{(M)} = \bm{N}_{p}^{(M)}$.
Note that the interaction radiation torque will appear only if the probe sphere is absorptive~\cite{silva:epl} and
the effective incident wave is asymmetric with respect to $O_p$.

\section{Numerical results and discussion }
We compute the radiation force and torque on each constituent of a system formed by two and three 
olive oil droplets suspended in water  at room temperature.
The choice of olive oil is because it is immiscible in water. 
The acoustic parameters of  water are $\rho_0=1000~\si{kg/m^3}$ and $c_0=1480~\si{m/s}$;
whereas for olive oil,  the parameters are $\rho_{q}=\unit915.8~\si{kg/m^3}, c_{q}=1464~\si{m/s}$, 
and $\alpha_{0,q}=4.10\times 10^{-14}~\si{Np/ (MHz^2~m)}$. 
Two types of external waves are considered, namely traveling and standing plane  waves 
at $1~\si{MHz}$ frequency. 
Unless mentioned otherwise, the droplets are considered in the Mie scattering regime
to which $ka_q\gtrsim 1$.

A \textsc{Matlab} (MathWorks Inc,) code was developed to  solve numerically the multiple scattering problem
represented by the linear system in (\ref{bnm2}). 
To test the multiple scattering code, we recovered the scattering form-function for  a tilted traveling plane wave
interacting with two spheres, as given  in Ref.~\cite{gaunaurd}.
But for the sake of brevity we will not present this result here.

The truncation order used to compute the effective beam-shape coefficient $n=M$ was determined
from the following condition
\begin{equation}
\label{trunc}
\left| \frac{b_{00,q}}{b_{Mm,q} }\right|< 10^{-6}, \quad q=1,2,\dots,N.
\end{equation}
This criterion was achieved for spheres with size parameter $ka_q=1$ by setting $M = 12$.

The results will be presented in terms of the dimensionless radiation force $\bm{Y}_p^{(M)}=\bm{F}^{(M)}_p/(E_0 a_p^2)$
and the dimensionless radiation torque $\bm{\tau}_p^{(M)}=\bm{N}^{(M)}_p/(E_0 a_p^3)$.

\subsection{Traveling plane wave}
Consider that an external plane wave  propagates along the $+z$-direction and interacts with spheres
placed at $\bm{r}'_q$ $(q=1,2,\dots,N)$.
The normalized velocity potential of the external wave with respect to $O_q$ 
(the system determined by the droplet at $\bm{r}'_q$)
is given by
\begin{equation}
\phi_\textrm{ex}(z)= e^{\ii k (z-z_q')}.
\end{equation}
The corresponding beam-shape coefficient is~\cite{colton}
\begin{equation}
\label{anm_plane}
a_{nm,q}= \ii^n \sqrt{4 \pi (2n+1)} e^{-\ii k z_q'} \delta_{m,0},
\end{equation}
where $\delta_{m,m'}$ is the Kronecker delta symbol.

\begin{figure}
\centering
\includegraphics[scale=.5]{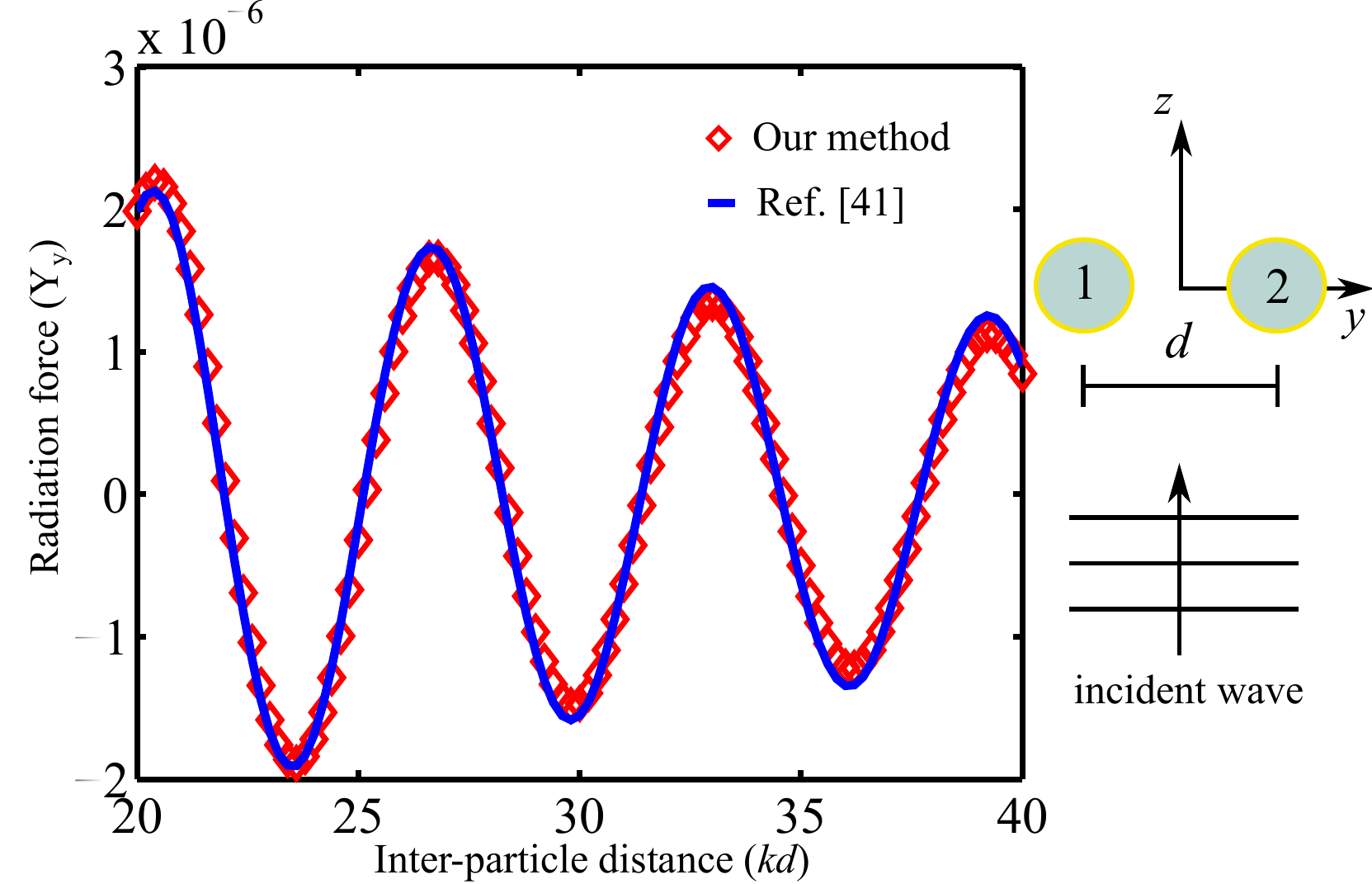}
\caption{(Color online) Acoustic interaction force between two rigid Rayleigh particles $(ka_1=ka_2=0.1)$  
induced by a plane wave propagating along the $+z$ direction.
The circles labeled as `$1$' and `$2$' denote, respectively, the particles located at $\bm{r}'_1=-(d/2)\bm{e}_y$
and $\bm{r}'_2=(d/2)\bm{e}_y$, where $d$ is the inter-droplet distance.
\label{fig2:validation}}
\end{figure}

The first example presented here is the case of the plane wave interaction with 
two rigid particles in the Rayleigh scattering limit, $ka_q=0.1$ $(q=1,2)$.
The particles are placed in the $y$-axis at $\bm{r}'_1=-(d/2)\bm{e}_y$
and $\bm{r}'_2=(d/2)\bm{e}_y$, where
$d$ is the inter-particle distance.
We want to compare the numerical result obtained with our method with
that derived by Zhuk~\cite{zhuk1985}, 
\begin{equation}
\label{zhuk}
 \bm{F}_{\textrm{int},1}=-\bm{F}_{\textrm{int},2} 
 = \frac{2 \pi E_0 k^3 a_1^3 a_2^3}{9}\frac{\sin k d}{kd} \bm{e}_y, \quad kd\gg1.
\end{equation}
The acoustic interaction force exerted on the sphere placed at $\bm{r}'_1$ is shown in Fig.~\ref{fig2:validation}. 
Excellent agreement is found between our method and the analytical result in (\ref{zhuk}).
The observed spatial oscillations in the interaction force is due to the stationary interference  pattern
of the re-scattered waves by each sphere.

\begin{figure}
\centering
\includegraphics[scale=.47]{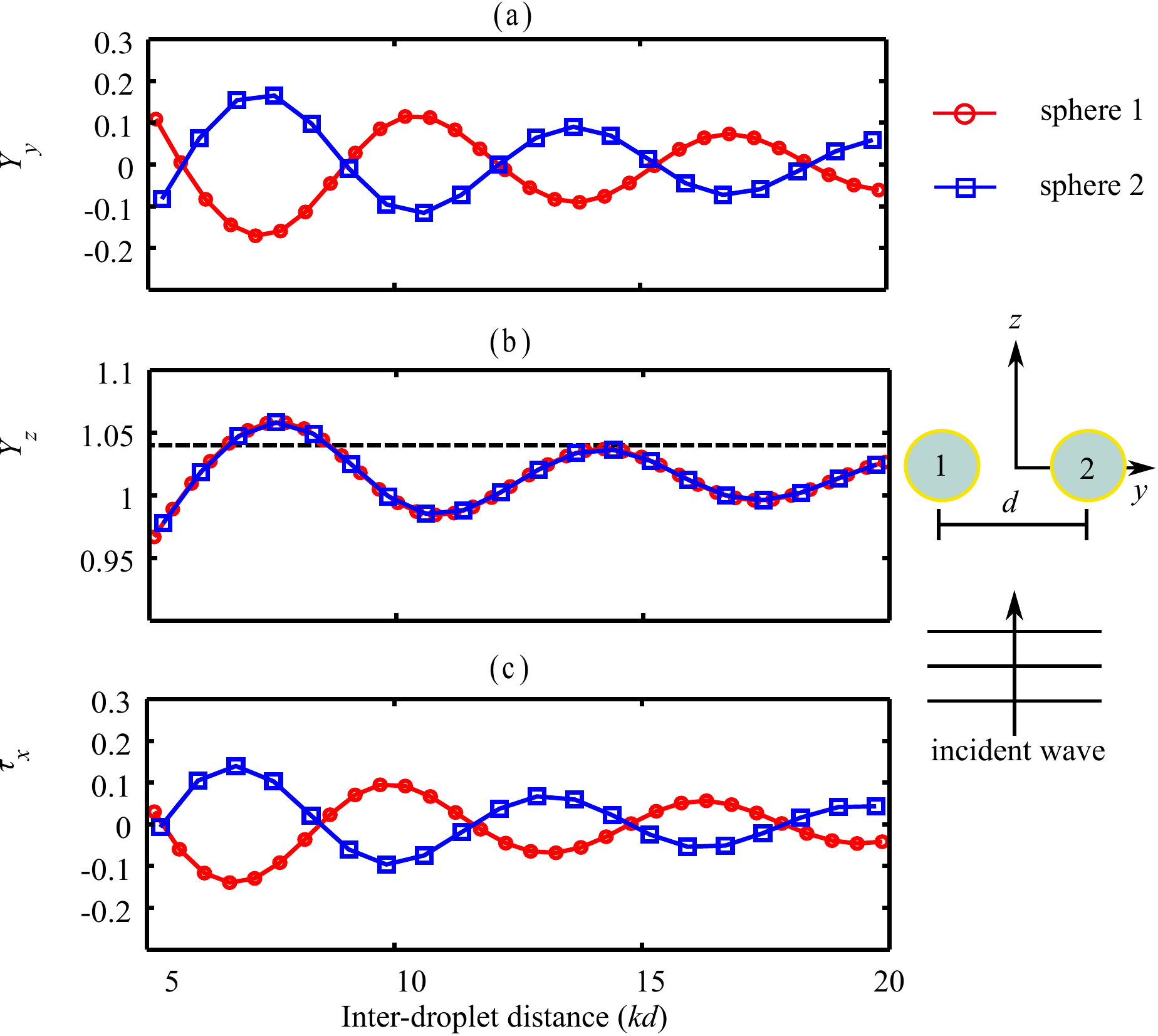}
\caption{(Color online)  
Acoustic interaction forces and torques between two olive oil droplets with size parameter $ka_1=ka_2=1$
induced by a plane wave propagating along the $+z$-direction.
The circles labeled as `$1$' and `$2$' denote, respectively, the droplets located at $\bm{r}'_1=-(d/2)\bm{e}_y$
and $\bm{r}'_2=(d/2)\bm{e}_y$, where $d$ is the inter-droplet distance.
The dashed line in (b) denotes the acoustic radiation force caused by the external traveling plane wave.
\label{fig3:RF_RT_PW_2_line}}
\end{figure}

In Fig.~\ref{fig3:RF_RT_PW_2_line}, we show the acoustic interaction forces and torques between two olive oil droplets with size parameter 
$ka_1=ka_2=1$.
The droplets are located at $\bm{r}'_1=-(d/2)\bm{e}_y$
and $\bm{r}'_2=(d/2)\bm{e}_y$, where $d$ is the inter-droplet distance.
Due to the symmetry between the droplets and the external wave, the interaction force along the $x$-axis is zero.
Likewise, using this symmetry argument, there is acoustic interaction torque only in the $x$-direction.
The interaction force in the $y$-direction is a pair antisymmetric force, i.e. 
$F^{(M)}_{\textrm{int},y,1} = - F^{(M)}_{\textrm{int},y,2}$. 
Here  the spheres interact directly to each other through the re-scattered waves.
In other words, the spatial amplitude variation of the external wave remains constant in the $xy$-plane at which
the spheres are positioned.
In contrast, the $z$-component of the interaction force is the same for each droplet with spatial oscillations
around the value of the radiation force due to the external wave (see the dashed line in Fig.~\ref{fig3:RF_RT_PW_2_line}.b).
This happens because the effective incident wave to both droplets, formed by the external and the re-scattered waves,
has the same interference pattern due to the symmetric position of the droplets.
Moreover, the $y$-component of the interaction force asymptotically approaches
zero as $d\rightarrow \infty$, whereas the $z$-component asymptotically approaches to
$1.04$, which corresponds to the radiation force on non-interacting droplets.
The acoustic interaction torque depends only on the direct interaction of the droplets through re-scattered waves.
Therefore, this interaction leads to a pair antisymmetric torque.
Moreover, the interaction torque also fades out as the droplets are set apart.

\begin{figure}
\centering
\includegraphics[scale=.45]{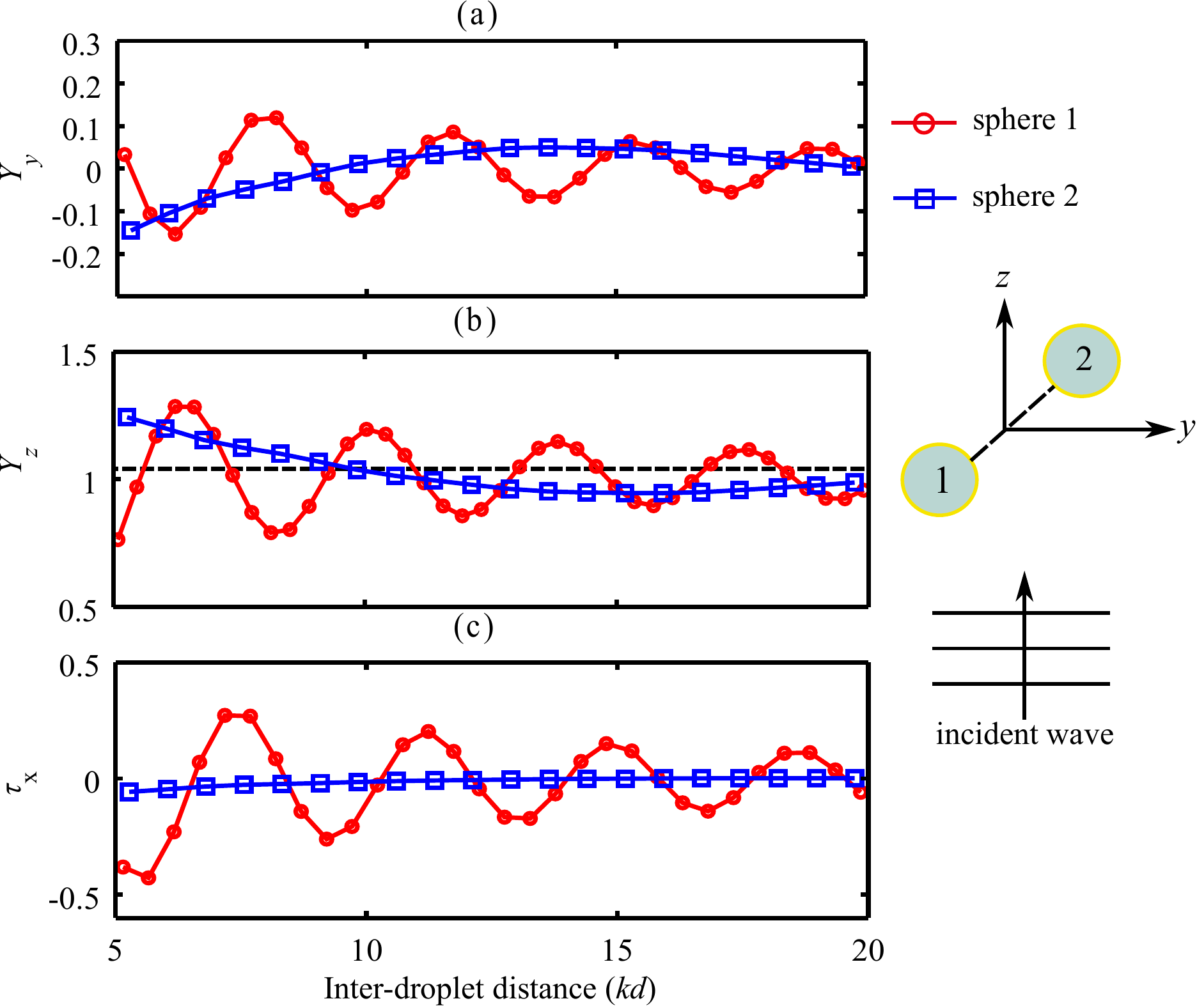}
\caption{(Color online) 
Acoustic interaction forces and torques between two olive oil droplets with size parameter $ka_1=ka_2=1$
 induced by a plane wave propagating along the $+z$-direction.
The circles labeled as `$1$' and `$2$' denote, respectively, the droplets located at
$\bm{r}'_1=-(d\sqrt{2}/2)(\bm{e}_x + \bm{e}_y)$
and $\bm{r}'_2=(d\sqrt{2}/2)(\bm{e}_x + \bm{e}_y)$, where $d$ is the inter-droplet distance.
The dashed line in (b) denotes the acoustic radiation force caused by the external traveling plane wave.
\label{fig4:RF_RT_PW_2_45}}
\end{figure}

The acoustic interaction forces and torques caused by the traveling plane wave  on  two olive oil droplets
with size parameters $ka_1=ka_2=1$ are shown Fig.~\ref{fig4:RF_RT_PW_2_45}
The droplets are positioned at $\bm{r}'_1=-(d\sqrt{2}/2)(\bm{e}_x + \bm{e}_y)$
and $\bm{r}'_2=(d\sqrt{2}/2)(\bm{e}_x + \bm{e}_y)$.
The axis that connects the droplets is tilted by $45^\circ$ with respect to the plane wave propagation direction.
No acoustic interaction force is present in the $x$-direction due to the symmetry of the 
droplets and the external wave.
Furthermore, the only component of the acoustic interaction torque is along the $x$-direction,
because  the effective incident wave to the droplets
lies on the $yz$-plane only.
The droplet located at $\bm{r}'_1$ interacts with an effective asymmetric stationary wave
formed by the external and re-scattered waves. 
This explains the oscillatory pattern seen on the acoustic interaction force and torque.
For the droplet at $\bm{r}'_2$ the external and re-scattered waves propagate almost
along the same direction.
This leads to a mild correction on the radiation force  due to the traveling plane wave.
We remark that a traveling plane wave does not produce torque on a single particle.
Note also that both interaction force and torque  asymptotically approach to zero
as the droplets are set apart.

\begin{figure}
\centering
\includegraphics[scale=.5]{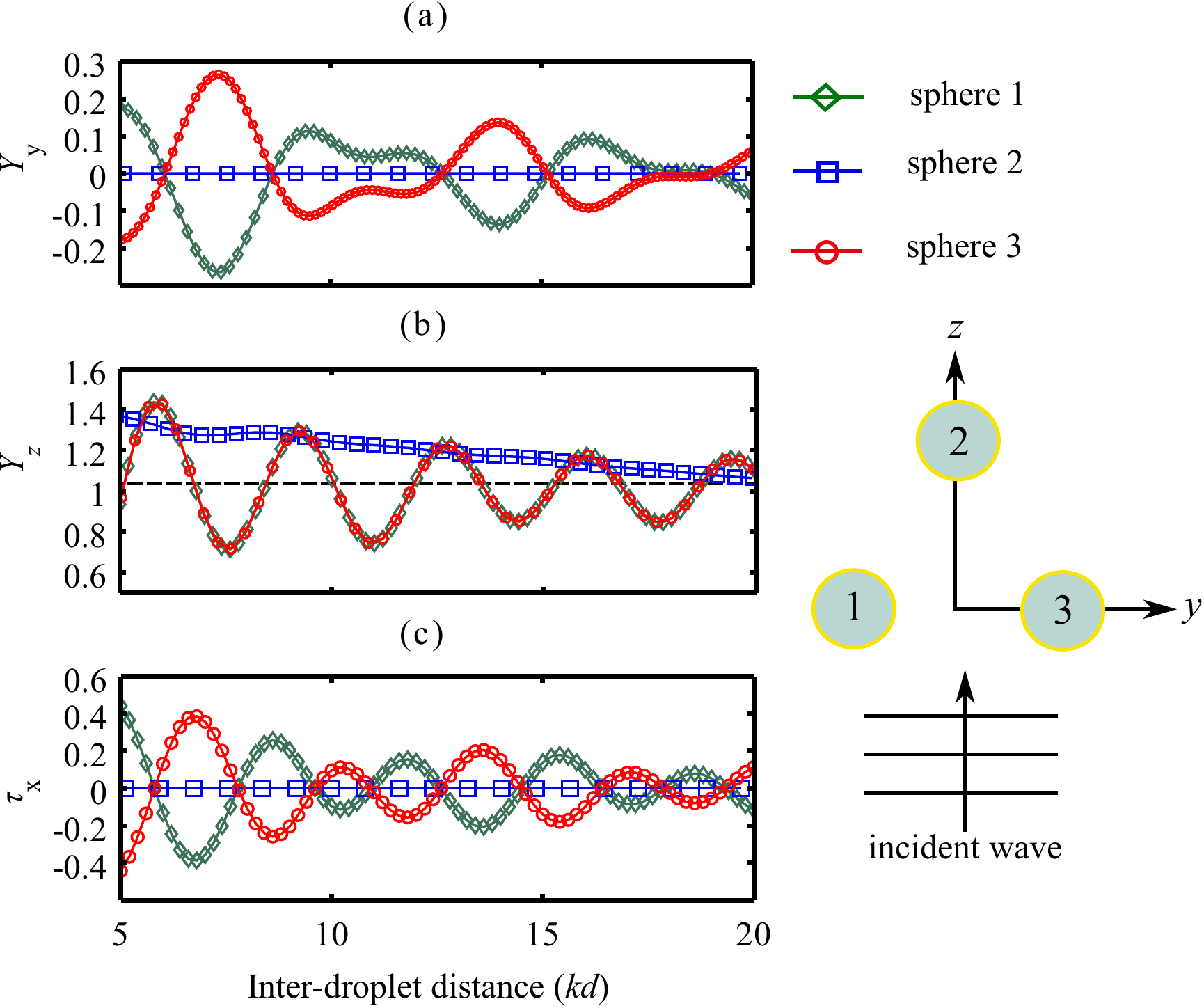}
\caption{(Color online) 
Acoustic interaction forces and torques between three olive oil droplets with size parameter $ka_1=ka_2=k a_3=1$ induced by a plane wave propagating along the $+z$-direction.
The circles labeled as `$1$', `$2$',  and `$3$' denote, respectively, the droplets located at $\bm{r}'_1=-(d/2)\bm{e}_y$,
$\bm{r}'_2=-(d \sqrt{3}/2)\bm{e}_z$, and
$\bm{r}'_3=(d/2)\bm{e}_y$, where $d$ is the inter-droplet distance.
The dashed line in (b) denotes the acoustic radiation force caused by the external traveling plane wave.
\label{fig:RF_RT_PW3}}
\end{figure}

In Fig.~\ref{fig:RF_RT_PW3}, we present the acoustic interaction force and torque exerted on three olive oil droplets
with size parameters $ka_1=ka_2=ka_3=1$.
The droplets are placed at  $\bm{r}'_1=-(d/2)\bm{e}_y$,
$\bm{r}'_2=-(d \sqrt{3}/2)\bm{e}_z$, and
$\bm{r}'_3=(d/2)\bm{e}_y$, where $d$ is the inter-droplet distance.
The effective incident wave to droplets at $\bm{r}'_1$ and $\bm{r}'_3$ is a stationary wave formed
by the external plane wave and the backscattered waves from the droplet at $\bm{r}'_2$.
Thus, the $z$-component of the interaction force are the same on the droplets at $\bm{r}'_1$ and $\bm{r}'_3$
due to their symmetrical position.
Moreover, the effective incident stationary wave  causes the oscillatory pattern in the interaction force.
The droplet located at $\bm{r}'_2$ experiences an effective incident wave composed by
the external plane wave and the forward scattered waves by the droplets at $\bm{r}'_1$ and $\bm{r}'_3$.
Hence, the $z$-component of the interaction force on the droplet at $\bm{r}'_2$ is larger than
the radiation force magnitude due to the external wave only, which values $1.04$.
The $y$-component of the interaction forces on the droplets at $\bm{r}'_1$ and $\bm{r}'_3$
form an antisymmetric force-pair because these droplets interact directly through  re-scattered waves
from each other.
Likewise the interaction torques  on these droplets also form an antisymmetric torque-pair.
Meanwhile, no interaction torque appears on the droplet at $\bm{r}'_2$ because of its symmetrical position
with respect to the other droplets and the external wave.

\begin{figure}
\centering
\includegraphics[scale=.6]{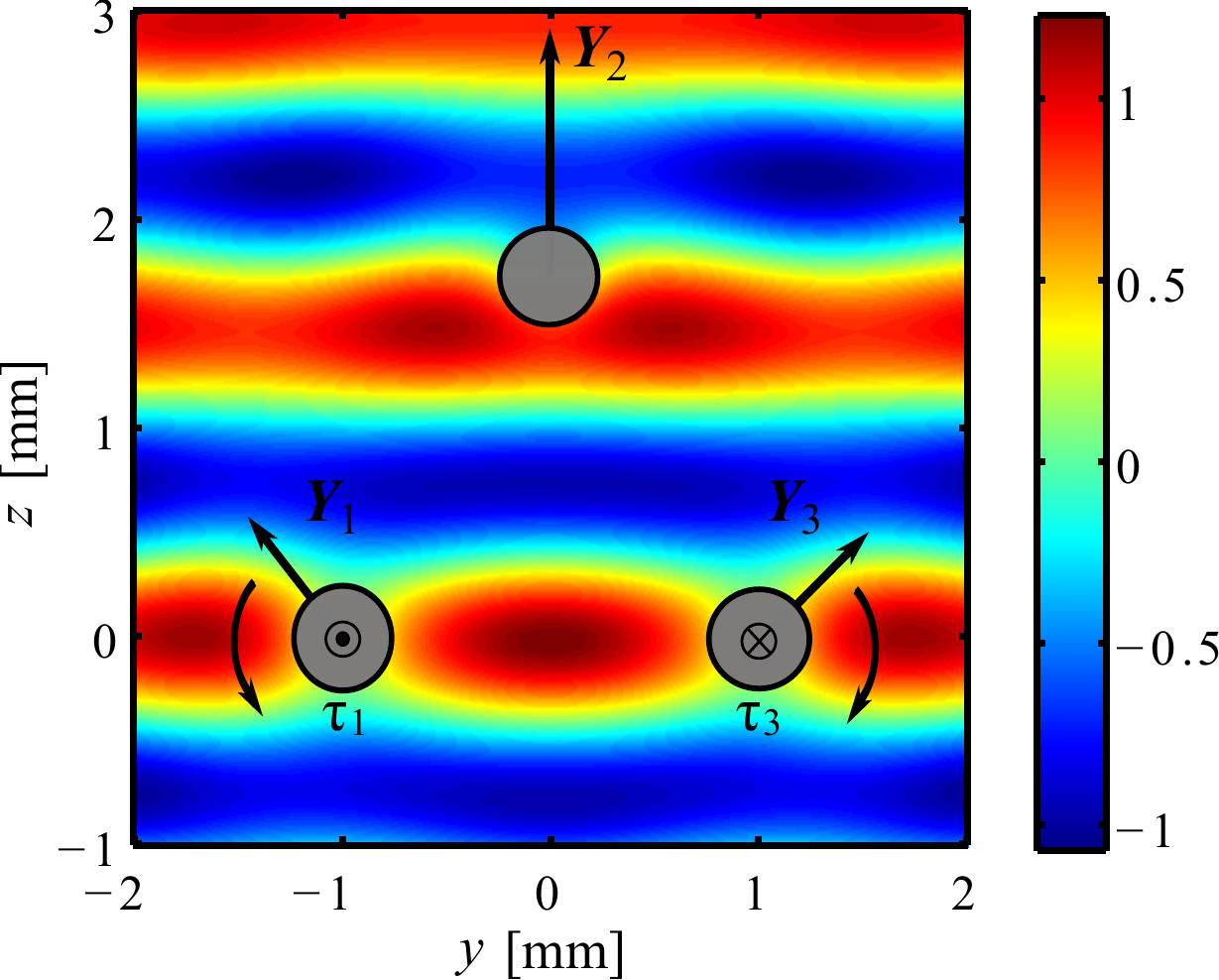}
\caption{(Color online) Acoustic interaction forces and torques exerted on three  olive oil droplets with size parameter $ka_1=ka_2=ka_3=1$ 
by a traveling plane wave along the $+z$-direction.
The droplets are located at $\bm{r}'_1=-(d/2)\bm{e}_y$,
$\bm{r}'_2=-(d \sqrt{3}/2)\bm{e}_z$, and
$\bm{r}'_3=(d/2)\bm{e}_y$, with the inter-droplet distance being $d=2~\si{mm}$.
The dimensionless acoustic forces, which are depicted by the straight arrows, 
are $\bm{Y}_1^{(M)}= -0.06\bm{e}_y + 1.01 \bm{e}_z$, $\bm{Y}_2^{(M)} = 1.28\bm{e}_z$, and $\bm{Y}_3^{(M)}= 0.06\bm{e}_y + 1.01 \bm{e}_z$.   
The dimensionless interaction torques are represented by $\odot$ (outward vector to the $yz$-plane)
and $\otimes$ (inward vector to the $yz$-plane) and they value $\tau_{1,x}^{(M)} = -\tau_{3,x}^{(M)} = 0.24$. 
The background is the amplitude of the external plus the scattered waves.
\label{fig:pw_plot_2D}}
\end{figure}

In Fig.~\ref{fig:pw_plot_2D},  we show the interaction of an external traveling plane wave with three olive oil droplets
with size parameters $ka_1=ka_2=ka_3=1$.
The background shows the amplitude of the external plus scattered waves.
The droplets are placed at $\bm{r}'_1=-(d/2)\bm{e}_y$,
$\bm{r}'_2=-(d \sqrt{3}/2)\bm{e}_z$, and
$\bm{r}'_3=(d/2)\bm{e}_y$, with the inter-droplet distance being $d=2~\si{mm}$.
The values of the dimensionless acoustic interaction force which arise on each droplet is $\bm{Y}_1^{(M)}= -0.06\bm{e}_y + 1.01 \bm{e}_z$,
$\bm{Y}_2^{(M)} = 1.28\bm{e}_z$, and $\bm{Y}_3^{(M)}= 0.06\bm{e}_y + 1.01 \bm{e}_z$.
The dimensionless interaction torque on the droplets at $\bm{r}'_1$ and $\bm{r}'_3$ is   
$\tau_{1,x}^{(M)} = -\tau_{3,x}^{(M)} = 0.24$.

\subsection{Standing plane wave}
The case of acoustic interaction forces and torques exerted on multiple spheres  in a standing plane   wave 
field is of great importance in acoustophoresis applications.
Assume that a standing  plane  wave is formed along the $z$-axis and
interacts with a system of spheres placed at $\bm{r}_q'$ $(q=1,2,\dots,N)$.
The amplitude of the velocity potential of the standing wave with respect to the system $O_q$
is given by
\begin{equation}
\phi_\textrm{ex}(z)=  \cos[ k (z-z_q')].
\end{equation}
Using (\ref{anm_plane}), we find that the corresponding beam-shape coefficient  of the standing wave is
\begin{equation}
a_{nm,q} =  \sqrt{4 \pi (2n+1)} \cos\left( k z_q'  - \frac{n\pi}{2} \right) \delta_{m,0}.
\end{equation}

\begin{figure}
\centering
\includegraphics[scale=.5]{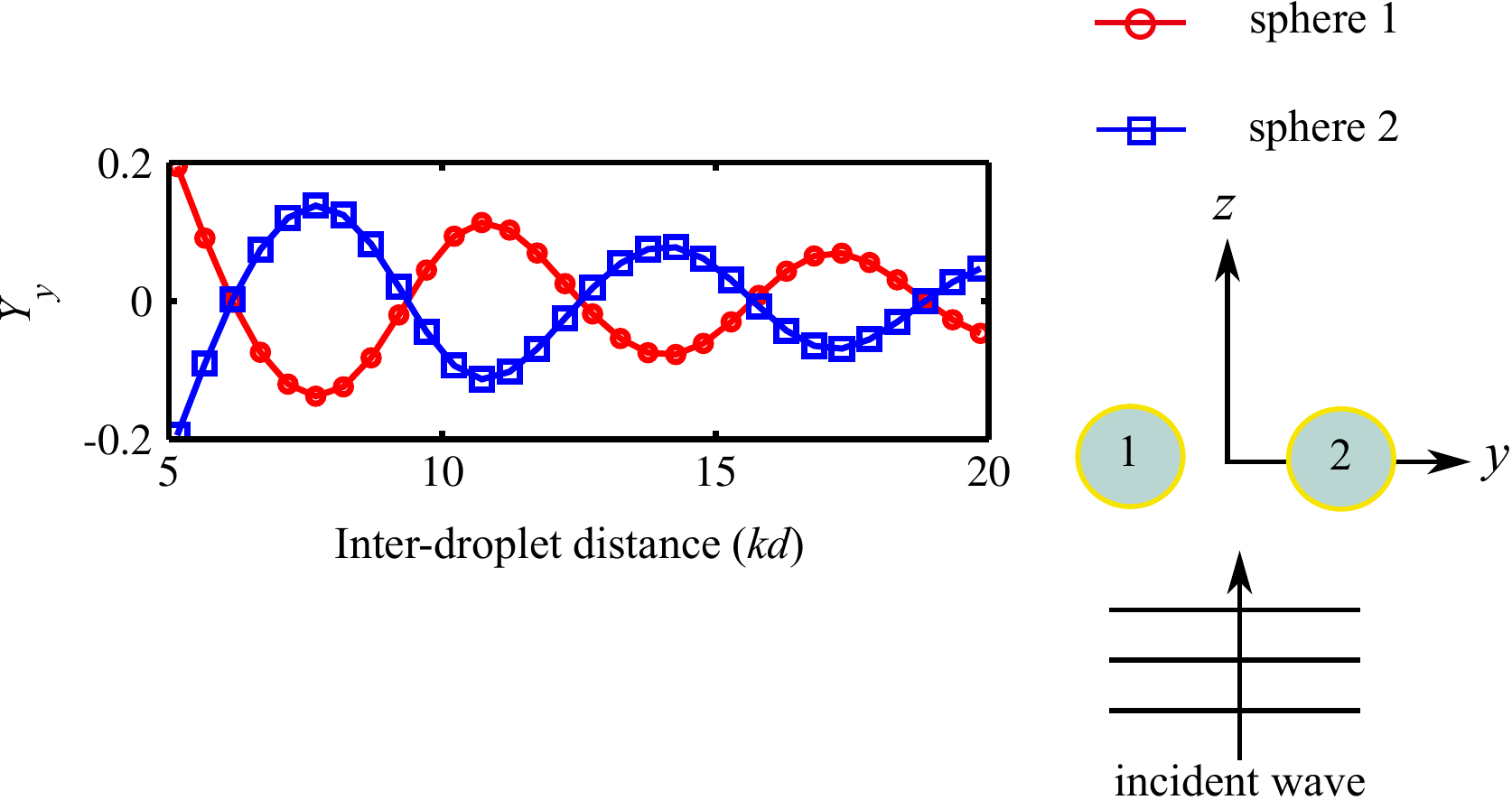}
\caption{(Color online) The $y$-component of the 
acoustic interaction forces between two olive oil droplets with size parameter $ka_1=ka_2=1$
induced by a standing plane wave along the $z$-direction.
The circles labeled as `$1$' and `$2$' denote, respectively, the droplets are located at $\bm{r}'_1=-(d/2)\bm{e}_y$
and $\bm{r}'_2=(d/2)\bm{e}_y$, where $d$ is the inter-droplet distance.
\label{fig7:RF_RT_SW_2_line}}
\end{figure}

The acoustic interaction forces exerted on two olive oil droplets with size parameter $ka_1=ka_2=1$ by
the standing wave is illustrated in Fig.~\ref{fig7:RF_RT_SW_2_line}.
The droplets  are located at $\bm{r}'_1=-(d/2)\bm{e}_y$
and $\bm{r}'_2=(d/2)\bm{e}_y$, where $d$ is the inter-droplet distance.
The droplets are in a node of the external standing wave $z=0$. 
Thus, the $z$-component of the radiation force on both droplets is zero~\cite{gorkov1962}.
Furthermore, no interaction torque is produced on the droplets because the effective incident wave
is symmetric with respect to the droplets.

\begin{figure}
\centering
\includegraphics[scale=.45]{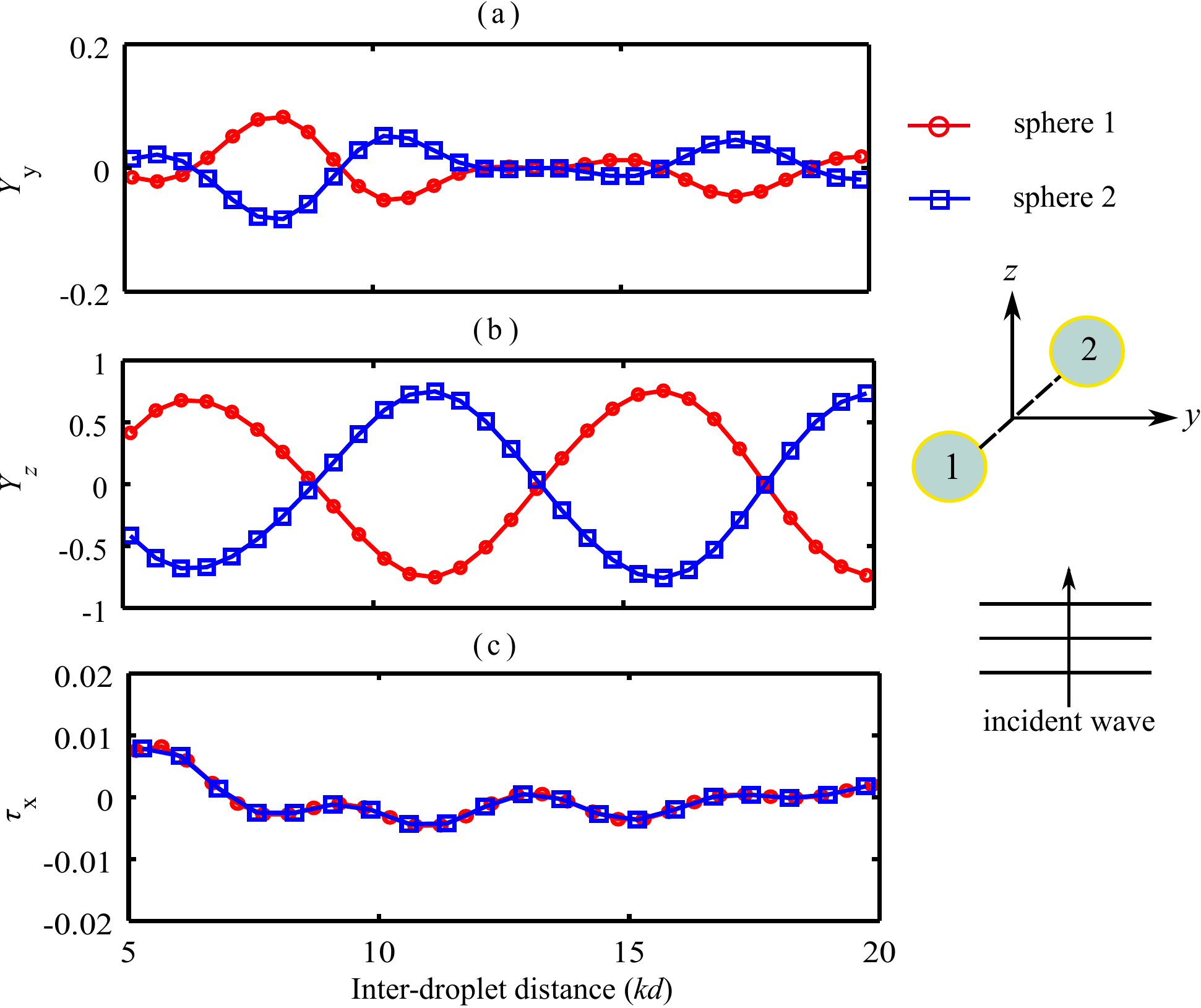}
\caption{(Color online) Acoustic interaction forces and torques between  
two olive oil droplets with size parameter $ka_1=ka_2=1$
induced by a standing plane wave along the $z$-direction.
The circles labeled as `$1$' and `$2$' denote, respectively, the droplets located at $\bm{r}'_1=\bm{0}$
and $\bm{r}'_2=(d\sqrt{2}/2)(\bm{e}_x + \bm{e}_y)$, where $d$ is the inter-droplet distance.
\label{fig8:RF_RT_SW_2_45}}
\end{figure}

In Fig.~\ref{fig8:RF_RT_SW_2_45}, we show the acoustic interaction forces and torques caused by 
the standing plane wave on  two olive oil droplets
with size parameters $ka_1=ka_2=ka_3=1$.
The droplets are placed at $\bm{r}'_1=-(d\sqrt{2}/2)(\bm{e}_x + \bm{e}_y)$
and $\bm{r}'_2=(d\sqrt{2}/2)(\bm{e}_x + \bm{e}_y)$.
Due to the symmetrical position of the droplets with respect to the external wave, 
no acoustic interaction force appears in the $x$-direction.
Moreover, the amplitude of the effective incident wave is the same on both droplets.
Thus, the acoustic interaction forces along the $y$-direction form an antisymmetric 
force-pair $F^{(M)}_{y,1} = -F^{(M)}_{y,2}$, because the effective incident wave
on a droplet is related to the backscattered wave of the opposite droplet. 
The $y$-component of the interaction force and  torque asymptotically approach to zero
as the droplets are set apart.
It is further noticed that the $z$-component of the force exerted
on the droplets is mostly caused by the external standing wave.
According to Gorkov's result~\cite{gorkov1962}, the radiation force varies spatially as $\sin(2kz)$.
Thus, this force is an odd function of $z$, which explains the antisymmetric pattern of the 
radiation force in the $z$-direction.
This and fact that the droplets interact through each other's backscattered wave, explain why  the  interaction torques 
on both spheres are equal.

\begin{figure}
\centering
\includegraphics[scale=.5]{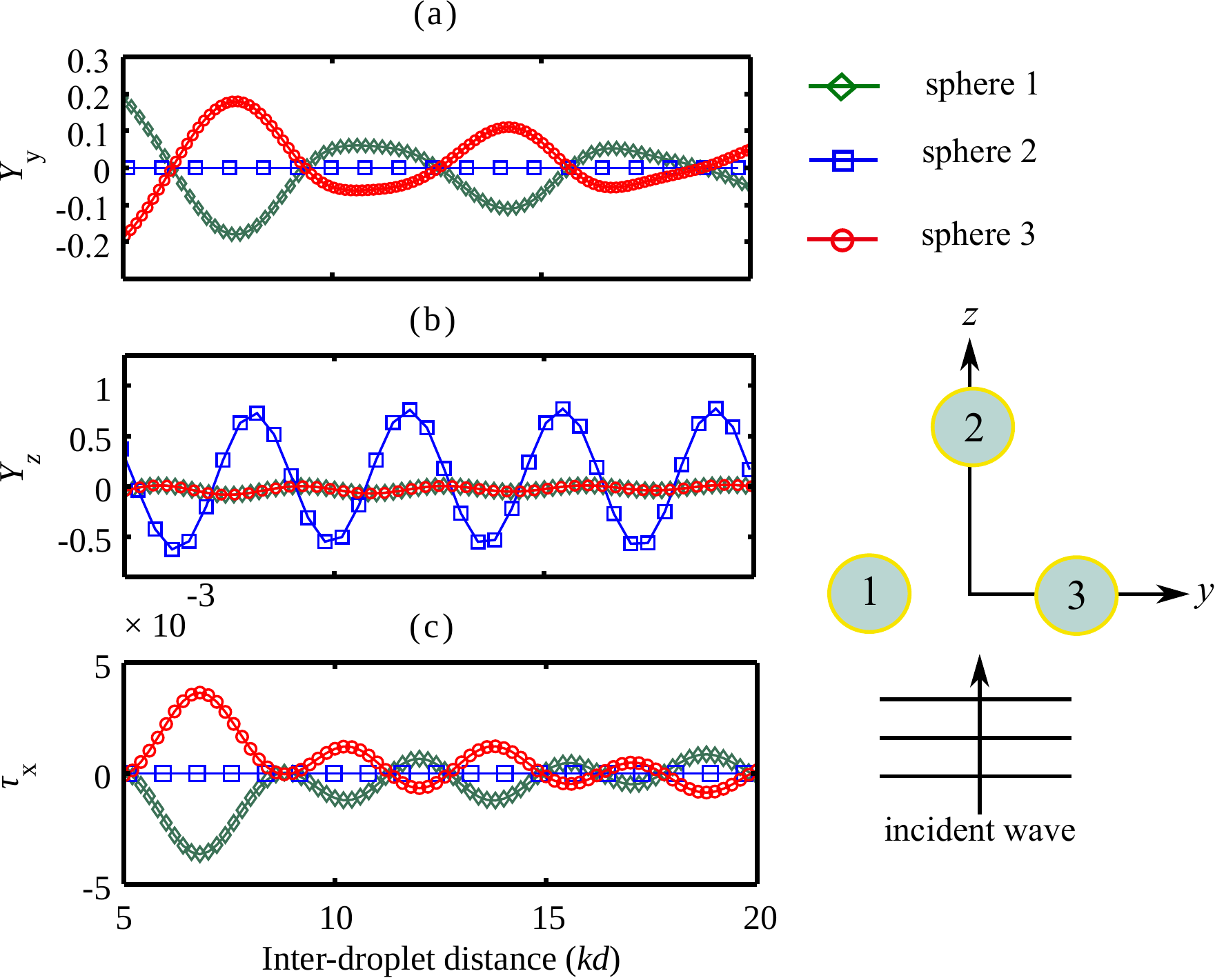}
\caption{(Color online) Acoustic interaction forces and torques between three olive oil droplets with size parameter $ka_1=ka_2=k a_3=1$ induced by a standing plane wave  along  the $z$-direction.
The circles labeled as `$1$', `$2$',  and `$3$' denote, respectively, the droplets located at $\bm{r}'_1=-(d/2)\bm{e}_y$,
$\bm{r}'_2=-(d \sqrt{3}/2)\bm{e}_z$, and
$\bm{r}'_3=(d/2)\bm{e}_y$, where $d$ is the inter-droplet distance.
\label{fig9:RF_RT_SW_3}}
\end{figure}

In Fig.~\ref{fig9:RF_RT_SW_3}, we show the acoustic interaction forces and torques exerted on three olive oil droplets ($ka_1=ka_2=ka_3=1$)
by the external standing plane wave.
The droplets are placed at  $\bm{r}'_1=-(d/2)\bm{e}_y$,
$\bm{r}'_2=-(d \sqrt{3}/2)\bm{e}_z$, and
$\bm{r}'_3=(d/2)\bm{e}_y$, where $d$ is the inter-droplet distance.
The  antisymmetric  force pair along the $y$-direction and
the interaction torque on the $x$-direction exerted on the droplets at $\bm{r}'_1$ and $\bm{r}'_3$ are formed because
the re-scattered of these droplets are equal but propagate in opposite directions.
On the other hand, the $z$-component of the radiation force is mostly due to the external
standing wave.
For the droplets located at $\bm{r}'_1$ and $\bm{r}'_3$, we have $z=0$; thus,
the radiation force along $z$-direction is nearly zero.
Whereas, this force varies as $\sin(2kz)$ on the droplet at $\bm{r}'_2$.

In Fig.~\ref{fig:pw_plot_2D},  we show the interaction of an external standing plane wave with three olive oil droplets
with size parameters $ka_1=ka_2=ka_3=1$.
The background is the amplitude of the external plus scattered waves.
The droplets are placed at $\bm{r}'_1=-(d/2)\bm{e}_y$,
$\bm{r}'_2=-(d \sqrt{3}/2)\bm{e}_z$, and
$\bm{r}'_3=(d/2)\bm{e}_y$, with the inter-droplet distance being $d=2~\si{mm}$.
The values of the dimensionless acoustic interaction force  on each droplet is 
$\bm{Y}_1^{(M)}= -0.13\bm{e}_y -0.3 \bm{e}_z$, $\bm{Y}_2^{(M)} = 0.63\bm{e}_z$, and $\bm{Y}_3^{(M)}= 0.13\bm{e}_y - 0.3 \bm{e}_z$
The dimensionless interaction torque on the droplets at $\bm{r}'_1$ and $\bm{r}'_3$ is   
$\tau_{1,x}^{(M)} = -\tau_{3,x}^{(M)} =0.003$.

\begin{figure}
\centering
\includegraphics[scale=.55]{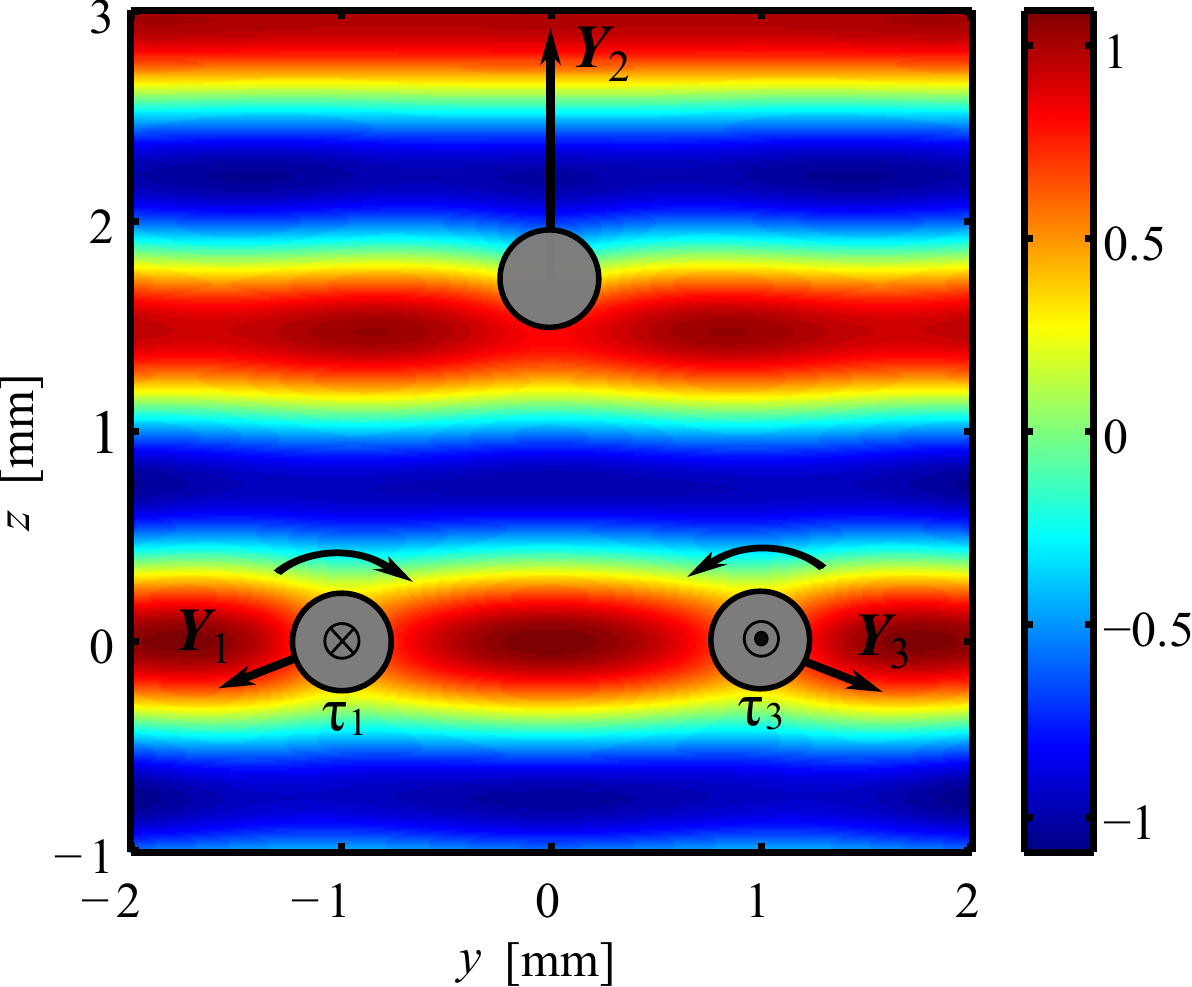}
\caption{(Color online) Acoustic interaction forces and torques exerted on three  olive oil droplets with size parameter $ka_1=ka_2=ka_3=1$ 
by a standing plane wave along $z$-direction.
The droplets are located at $\bm{r}'_1=-(d/2)\bm{e}_y$,
$\bm{r}'_2=-(d \sqrt{3}/2)\bm{e}_z$, and
$\bm{r}'_3=(d/2)\bm{e}_y$, with the inter-droplet distance being $d=2~\si{mm}$.
The dimensionless acoustic forces, which are depicted by the straight arrows, 
are $\bm{Y}_1^{(M)}= -0.13\bm{e}_y -0.3 \bm{e}_z$, $\bm{Y}_2^{(M)} = 0.63\bm{e}_z$, and $\bm{Y}_3^{(M)}= 0.13\bm{e}_y - 0.3 \bm{e}_z$.
The dimensionless interaction torques are represented by $\odot$ (outward vector to the $yz$-plane)
and $\otimes$ (inward vector to the $yz$-plane) and they value $\tau_{1,x}^{(M)} = -\tau_{3,x}^{(M)} = 0.003$. 
The background is the amplitude of the external plus the scattered waves.
\label{fig10:RF_RT_SW_image}}
\end{figure}

\section{Conclusion}
We have presented a method to compute the acoustic interaction forces and torques on a cluster of $N$ spheres 
suspended in an inviscid fluid. 
The  method is based on the solution of the corresponding multiple scattering problem using  partial-wave expansions
of spherical functions.
With this solution, the radiation torques and forces exerted on the spheres are calculated using the farfield approach
developed, respectively, in~\cite{silva:epl,silva:3541}.

Moreover, we have analyzed the  interaction forces and torques on arrangements  
of two and three olive oil droplets (fluid compressible spheres) with size parameter $ka_q=1$, which corresponds 
to the so-called Mie sized particles.
We emphasize also that by computing the scaled scattering coefficient $s_{q,n}$, 
the proposed method can be readily extended to  solid elastic and
viscoelastic, and layered materials. 
In particular, both traveling and standing plane waves were used as the external waves to the droplets.
The acoustic interaction force is highly dependent to the relative position between the droplets.
More specifically, they depend on the interference pattern formed between the external and re-scattered waves.
Additionally, we have shown that the acoustic interaction torque may appear on the droplets
depending on the asymmetry between the effective incident wave (external plus re-scattered waves) 
and droplets.

The proposed method can be useful to understand  many-particle dynamics under an ultrasound
wave in experiments performed in acoustofluidics and acoustical tweezer devices.

\appendix
\section{Expansion coefficients}
Using the orthonormality property of the spherical harmonics in (\ref{phi_ext}) and (\ref{snm_qp}),
we  obtain the beam-shape and scattering coefficient with respect to the system $O_p$ as
\begin{align}
\nonumber
\left[
\begin{matrix}
 a_{nm,p} \\
 s_{nm,qp}
\end{matrix}
\right] &= 
\left[
\begin{matrix}
 j^{-1}_n(k R_p) \\
 h_n^{(1)}\mbox{}^{-1}(k R_p)
\end{matrix}
\right]
\int_{0}^\pi \int_{0}^{2\pi} \left[
\begin{matrix}
 \phi_{\text{ex},p}(kR_p,\theta_p,\varphi_p) \\
 \phi_{\text{sc},qp}(kR_p,\theta_p,\varphi_p)
\end{matrix}
\right]\\ 
&\times Y_{n}^{m*}(\theta_p,\varphi_p) \sin \theta
\dd\theta \dd\varphi,
\label{bsapp}
\end{align}
where $R_p$ is the radius of any spherical region centered at $O_p$ enclosing the particle at $\bm{r}'_p$.
The incident and the scattered waves can be expanded in the form of a partial-series with respect to another system, $O_q$, as
\begin{equation}
\label{pi2}
 \left[
\begin{matrix}
 \phi_{\text{ex},q}(\bm{r}_q) \\
 \phi_{\text{sc},qp}(\bm{r}_q)
\end{matrix}
\right] = \sum_{\nu,\mu} 
\left[
\begin{matrix}
a_{\nu\mu,q} J_\nu^\mu (\bm{r}_q) \\
s_{\nu\mu,qp} H_\nu^\mu (\bm{r}_q) \\
\end{matrix}
\right].
\end{equation}

The addition theorem provides a way to express the partial-waves in the $O_q$ in terms of any other coordinate system, say $O_p$.
Accordingly,~\cite{ivanov}
\begin{equation}
\left[
\begin{matrix}
J_\nu^\mu (\bm{r}_q) \\
H_\nu^\mu (\bm{r}_q) \\
\end{matrix}
\right] = \sum_{\nu,\mu} 
\left[
\begin{matrix}
S_{n\nu}^{m\mu,1}(\textit{\textbf{r}}_{qp}') \\
S_{n\nu}^{m\mu,2}(\textit{\textbf{r}}_{qp}') \\
\end{matrix}
\right] J_\nu^\mu( \bm{r}_{p}),
\label{addition}
\end{equation}
where  $\textit{\textbf{r}}_{qp}' = \textit{\textbf{r}}_{q}'- 
\textit{\textbf{r}}_{p}' = \textit{\textbf{r}}_{q}- \textit{\textbf{r}}_{p} $ 
is the vector from  $O_q$ to  $O_p$.
The translation coefficients are
\begin{align}
\nonumber
\left[
\begin{matrix}
S_{n\nu}^{m\mu,1}(\textit{\textbf{r}}_{qp}') \\
S_{n\nu}^{m\mu,2}(\textit{\textbf{r}}_{qp}') \\
\end{matrix}
\right]  &= 4\pi \ii^{\nu-n} \sum_{\sigma=|{n-\nu}|}^{n+\nu}
 \ii^{\sigma}(-1)^{m} \mathcal{G}(n,m;\nu,\mu;\sigma) \\
&\times \left[
\begin{matrix}
J_{\sigma}^{\mu-m,l}(k r_{q},\theta_{qp},\phi_{qp}) \\
H_{\sigma}^{\mu-m,l}(k r_{q},\theta_{qp},\phi_{qp})\\
\end{matrix}
\right],
\label{addition1}
\end{align}
where $\theta_{qp}$ and $\varphi_{qp}$ are the polar and azimuthal angles of 
the vector
$\textit{\textbf{r}}_{qp}'$, and $\mathcal{G}(n,m;\nu,\mu;\sigma)$ is the 
Gaunt coefficient given in Ref.~\cite{ivanov}.
Now, substituting~(\ref{addition}) into~(\ref{phi_ext}) and (\ref{snm_qp}) and using the result
in~(\ref{bsapp}), one obtains 
\begin{equation}
\left[
\begin{matrix}
 a_{nm,q} \\
 s_{nm,qp}
\end{matrix}
\right] 
 =\sum_{\nu,\mu} \left[
\begin{matrix}
 a_{\nu \mu,p} S_{n\nu}^{m\mu,1}(\textit{\textbf{r}}_{qp}') \\
 s_{\nu \mu,q} S_{n\nu}^{m\mu,2} (\textit{\textbf{r}}_{qp}')
\end{matrix}
\right].
\label{app:bs_sc}
\end{equation}

\section*{\bf{Acknowledgments}}
This work partially was supported by Grant 303783/2013-3 CNPq (Brazilian agency).
M. Azerpeyvand would like to thank the financial support of the Royal Academy of Engineering.



\end{document}